\documentclass[mathleft
]{an}
\usepackage{graphicx}
\usepackage{times}
\usepackage{natbib}
\bibpunct{(}{)}{;}{a}{}{}

\def\cm-2{cm$^{-2}$}

\def\swift{{\it Swift}}
\def\chandra{{\it Chandra}}

\def\xmm{{XMM-Newton}}
\def\m31{\object{M~31}}
\def\me33{\object{M~33}}

\newcommand{\expo}[1]{$\times 10^{#1}$}

\def\n12b{M31N~2007-12b}

\begin{document}

\Pagespan{789}{}
\Yearpublication{2006}%
\Yearsubmission{2005}%
\Month{11}%
\Volume{999}%
\Issue{88}%

\title{X-ray emission from optical novae in M 31\,\thanks{Partly 
   based on observations with XMM-Newton, an ESA Science Mission 
    with instruments and contributions directly funded by ESA Member
    States and NASA}}

\author{W. Pietsch\,\thanks{Corresponding author:
  \email{wnp@mpe.mpg.de}\newline}
}
\titlerunning{X-ray emission from optical novae in M 31}
\authorrunning{W. Pietsch}
\institute{
Max-Planck-Institut f\"ur extraterrestrische Physik, Giessenbachstra\ss e, 
D-85748 Garching, Germany
}

\received{??}
\accepted{??}
\publonline{later}


\abstract{%
The first supersoft source (SSS) identification with an optical nova in \m31 
was based on ROSAT observations. Twenty additional X-ray counterparts (mostly
identified as SSS by their hardness ratios) were detected using archival ROSAT,
\xmm\  and \chandra\  observations obtained before July 2002. Based on these
results optical novae seem to constitute the major class of SSS in \m31. 
An analysis of archival 
\chandra\  HRC-I and ACIS-I observations obtained from July 2004 to February 2005
demonstrated that \m31  nova SSS states lasted from months to about 10 years.
Several novae showed short X-ray outbursts starting within 50 d after the
optical outburst and lasting only two to three months. The fraction of novae
detected in soft X-rays within a year after the optical outburst was more than 
30\%. Ongoing optical nova monitoring programs, optical spectral follow-up and 
an up-to-date nova catalogue are essential for the X-ray work. Re-analysis of
archival nova data to improve positions and find additional nova candidates are
urgently needed for secure recurrent nova identifications. 
Dedicated \xmm/\chandra\  monitoring programs for X-ray emission from
optical novae covering the center area of \m31  continue to provide interesting new results
(e.g. coherent 1105 s pulsations in the SSS counterpart of nova M31N 2007-12b). The 
SSS light curves of novae allow us -- together with optical information -- to
estimate the mass of the white dwarf, of the ejecta and the burned mass in the
outburst. Observations of the central area of \m31\ allow us -- in contrast to
observations in the Galaxy -- to monitor many novae simultaneously and proved to
be prone to find many interesting SSS and nova types.}

\maketitle

\section{Introduction}
The outbursts of classical novae (CNe) are caused by explosive hydrogen burning
on the white dwarf (WD) surface of a cataclysmic variable, a close binary
system with transfer of material from a main sequence star to the WD.
After about $10^{-7}-5\times10^{-4}$~M$_{\odot}$ of H-rich material are 
transferred to the WD, 
ignition under degenerate conditions takes place in the accreted
envelope and a thermonuclear runaway is initiated 
\citep[see e.g.][]{1998ApJ...494..680J,2005ApJ...623..398Y}. 
As a consequence, the envelope expands and causes the
brightness of the star to increase to
maximum luminosities of up to $\sim 10^{5}$~L$_{\odot}$. A fraction
of the envelope is ejected, 
while a part of it remains in steady nuclear burning
on the WD surface. This powers a supersoft X-ray source (SSS)
which can be observed as soon as the expanding ejected envelope becomes 
optically thin to soft X-rays, with the spectrum of a hot ($T_{eff}:
10^{5}-10^{6}$~K) WD atmosphere 
\citep[][]{1991ApJ...373L..51M}.
The duration of the SSS phase is inversely related to the 
WD mass while the time of appearance of the SSS is determined by
the mass ejected in the outburst and the ejection velocity.
Models of the post-outburst WD envelope show that  
steady H-burning can only occur for envelope masses smaller than 
$\sim 10^{-5}$~M$_{\odot}$ 
\citep[][]{2005A&A...439.1061S,1998ApJ...503..381T},
and the observed evolution
of the SSS in V1974 Cyg has been successfully modeled by an envelope of 
$\sim 2\times10^{-6}$~M$_{\odot}$ 
\citep[][]{2005A&A...439.1057S}.
WD envelope models show that the duration of the SSS state also depends 
on the metalicity of the envelope, so the monitoring of the SSS states of CNe 
provides constraints also on the chemical composition of the post-outburst 
envelope. 
\citet[][]{2006ApJS..167...59H} have
developed envelope and wind models that simulate the optical and X-ray light
curves for several WD masses and chemical compositions. 

Accreting WDs in recurrent novae (RNe) are good candidates for type
Ia supernovae (SNe) as RNe are believed to contain massive WDs.
However, one of the main
drawbacks to make RNe convincing progenitors of SNe-Ia was their low fraction
in optical surveys 
\citep[][]{1994ApJ...423L..31D}.
In the case of CNe the ejection of
material in the outburst makes it difficult to follow the
long-term evolution of the WD mass. For some CNe there
is a disagreement between theory and observations regarding the ejected
masses, with observational determinations of the mass in the ejected
shell larger than predicted by models. The duration of the SSS phase 
provides the only direct 
indicator of the post-outburst envelope mass remaining on the WD in RNe and CNe. In the case of
CNe with massive WDs, the SSS state is very short ($<$100 d) and could have been
easily missed in previous surveys. 
CNe with short SSS state are additional good candidates for SNe-Ia progenitors 
which makes determining their frequency very important.

Nevertheless, the number and duration of SSS states observed in optical novae 
are small: in a systematic search of X-ray emission from CNe in the ROSAT 
archive
\citep[][]{2001A&A...373..542O},
found only three novae with SSS emission in X-rays from a total of
39 CNe observed less than ten years after the outburst with SSS phases lasting
between 400~d and 9~yr.
The \chandra\ and \xmm\ observatories
have detected SSS emission for several more novae; but only a limited number 
of observations have been performed for each source, 
providing little constraints on the duration of the SSS state. Of specific
interest was
the monitoring of the recurrent nova RS~Oph in spring 2006 with
the Swift satellite which clearly determined the end of the SSS state less than
100~d after outburst 
\citep[see e.g.][]{2006ATel..838....1O}
which suggests a WD mass of 1.35 M$_{\odot}$ 
\citep[][]{2006ApJ...651L.141H}.
The observations of the Galactic nova V458 Vul 
\citep[detected as
highly variable SSS $\sim$400 d after outburst, see e.g.][]{2008ATel.1721....1D}
and nova V2491 Cyg 
\citep[starting its SSS state 36~d after the optical outburst, 
see e.g.][]{2008ATel.1573....1N}
demonstrated that each Galactic
nova seems to have its own peculiarities.

The small number of novae found to exhibit a SSS state,
and the diversity of the duration of this state (from 10 years down
to few weeks) present one of the big mysteries in the study of 
hydrogen burning objects over the last years. Despite an extensive
target of opportunity program with \chandra\ and \xmm\ (of order 3 dozen observations
during the last 4 years), little progress has been made in
constraining the duration of SSS states, or to even
putting constraints on the long term evolution of accreting 
WDs in binary systems. This now may change with the monitoring campaign for
Galactic novae with the \swift\ satellite (see Osborne et al. in this issue).

\begin{figure}
\includegraphics[width=82mm,clip=]{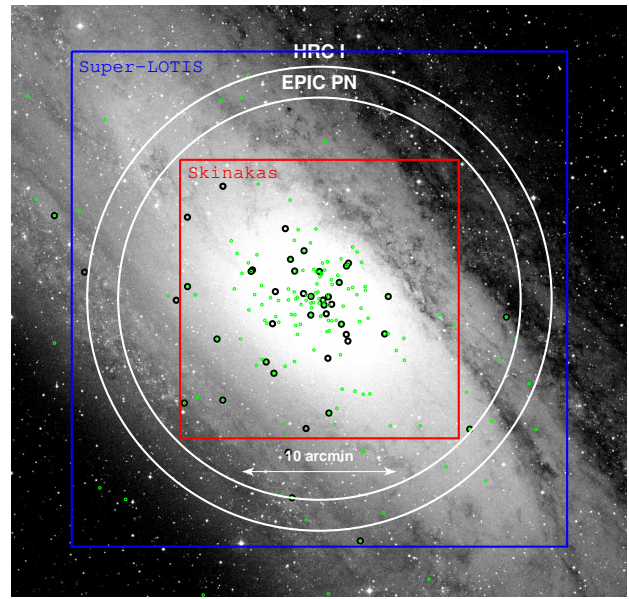}
\caption{DSS1 image of M\,31: the SuperLOTIS and Skinakas  
     field of view (FoV) are overlaid as big squares. Circles show the \chandra\ 
     HRC~I and \xmm\ EPIC PN fields. Novae with outbursts after 2000 
     are indicated by green dots. Big circles indicate X-ray detected novae (extracted
     from our \m31\ nova web page, see http://www.mpe.mpg.de/$\sim$m31novae/opt/m31/M31\_table.html).}
\label{fovima}
\end{figure}

\section{Why observe optical novae in \m31?}
In contrast to the Galaxy, observing novae in the nearby galaxy \m31\ offers the unique
chance to learn more about the duration of the supersoft
X-ray state in a global population of novae with minimal effort:

\begin{itemize}
\item[--]
\m31\ is the only nearby galaxy with many (more than 100 novae detected
over the last 5 years!) reported optical novae within the FoV of one \xmm\ or\linebreak 
\chandra\ observation (Fig.~\ref{fovima}).
\item[--]
All novae are effectively at the same, known distance, thus allowing
 easy comparison of light curves and maximum brightness/luminosity in the optical
 and X-ray\linebreak regime.
\item[--]
Low Galactic foreground as well as low intrinsic \m31\ absorption 
 allows accurate determination of color and temperature.
\item[--]
With comparatively little observing time one can obtain a homogeneous
 nova sample and follow the X-ray evolution of all of them
 over time.
\item[--]
With the derived time of appearance of the SSS and the durations of nova SSS phases one can constrain 
envelope and WD masses, and potentially correlate differences in metalicity
with location in \m31\ (bulge versus disk population).
\item[--]
X-ray parameters can be correlated with nova type\linebreak (Fe~II, He/N) and 
speed class for many novae.
\end{itemize}

\section{X-ray detection of optical novae in \m31} \label{Xnova}
Already in the ROSAT proposal for a ``Deep Survey of the
Andromeda Nebula" with the PSPC in 1989, it was speculated that several
tens of novae should be detectable in the survey  
\citep[extrapolating from EXOSAT detections of optical novae as SSSs;][]{1987A&A...177..110O}.
ROSAT has observed the full disk of the \m31\ galaxy twice (about 6.5 deg$^2$).
A ROSAT PSPC mosaic of 6 contiguous pointings with an exposure time of
25 ks each was performed in July 1991 
\citep[first \m31\ survey;][hereafter SHP97]{1997A&A...317..328S}.
A second survey
was made in July/August 1992 and January and July 1993  
\citep[][hereafter SHL2001]{2001A&A...373...63S}. Several SSS candidates were
identified. In a search for counterparts, 
only one recent nova (which erupted in 1990 in \m31) was found on optical
images \citep{2002A&A...389..439N}. 
The population of SSS in \m31\ has been studied by
\citet{2004ApJ...610..261G, 1996LNP96.472...75G}, in particular their variability.
One of the surprising results was that more fading than
rising sources have been found. Coincidentally, one of
these faders was the above-mentioned nova 
\citep[RX~J0044.0+4118;][]{2002A&A...389..439N}. 
This led to speculation that the
difference in the numbers of faders and risers is due to a fraction
of classical novae for which the X-ray rising phase could be
much shorter than the fading phase.
Based on the, until then, known durations of the supersoft
X-ray phases, this explanation was considered unlikely.
Also, the global (bulge+disk) nova rate of
$\sim$37 nova per year in \m31\ \citep{2001ApJ...563..749S},
combined with the short duration of the ROSAT survey,
did not suggest more than two novae among the two dozen ROSAT
SSSs in \m31\ when taking into account
the wide spread locations of the SSSs over the \m31\ disk.
Similarly, recurrent novae were not expected to contribute
to the observed SSS sample, since
the outburst rate of recurrent novae in \m31\ has been estimated to be only
10\% of the rate of classical novae \citep{1996ApJ...473..240D}. 

However, the very incomplete \m31\ optical nova catalogues at the time
were the most
likely reason for the low number of ROSAT SSS identifications.
Another reason may be that the limited spatial resolution of the 
ROSAT PSPC led to a reduced sensitivity for SSS detections in the crowded central 
\m31\ field where many of the optical novae 
are detected.

This situation changed with the advent of \chandra\ and \xmm\ and the improved
optical nova catalogues. 
In a search for X-ray emission from optical novae, 
\citet[][hereafter PFF2005]{2005A&A...442..879P}
correlated \m31\ X-ray source catalogues sensitive to the detection of SSS 
(ROSAT PSPC: SHP97, SHL2001; 
\chandra\ HRC-I: 
\citet[][hereafter K2002]{2002ApJ...578..114K};
\chandra\ ACIS-S: 
\citet[][hereafter DKG2004]{2004ApJ...610..247D};
\xmm: 
\citet[][hereafter PFH2005]{2005A&A...434..483P})
with the nova list of the Wendelstein Calar Alto Pixellensing Project 
\citep[WeCAPP,][]{2001A&A...379..362R} and novae from the 
literature. 
Within the ROSAT\linebreak lists, PFF2005 identified five novae.  
Within the \chandra\ HRC-I catalogues, eight nova counterparts were detected 
with count rates of 
(3--30)\expo{-4}~ct~s$^{-1}$. 
Five of these
novae have also been detected in ACIS-S 
and were
classified as SSS.  The \xmm\ observations of the \m31\ center (four pointings
with 6 month spacing  from June 2000 to January 2002)
revealed three additional novae. Four novae were
detected as SSS within a year after outburst. While nova N2000-03 turned on
as a SSS $\sim$180~d after outburst and was still active more than a 
year later, nova
N2001-06 was just detected once $\sim$120~d after outburst and
65 days later had dropped in X-ray flux by at least a factor of 10. 
The sample at the time more than tripled the number of known optical novae with
a SSS phase. The SSS phase of at least 15\% of the novae started within a year.   

\citet[][hereafter PHS2007]{2007A&A...465..375P} searched for optical nova 
counterparts in \chandra\ HRC-I and 
ACIS-I X-ray monitoring observations of the \m31\ 
center that were performed  from June 2004 to February 2005 to detect black hole
X-ray transients and time
variability similar to that from the Galactic center black hole (Sag A$^*$) also 
from the \m31\ center black hole (\m31$^*$). 
PHS2007 serendipitously
detected eleven out of 34 novae within a year after the optical outburst. 
While for eleven novae from PFF2005 they detect the end of the SSS phase, 
seven novae were still bright 1200 to 3380 days after the outburst. 
Several of the X-ray outbursts lasted less than 100 days. They found that
the number of optical novae detected as SSSs is much higher than
previously estimated ($>30$\%). From the X-ray light curves estimates for the
mass of the ejecta and of the mass burned were given.

\citet{2006IBVS.5720....1S} 
showed that one SSS candidate of PFH2005 correlated with the position of a 
nova which was detected on their optical
images 84 days before the X-ray detection. The position of this nova differs
from that of the nova close-by proposed as identification for the SSS by 
PFF2005 and is the more likely counterpart. 
Inspired by the success, \citet{2006IBVS.5737....1S}  
identified another up to then undiscovered optical nova with a PFH2005 SSS.

In a search for \m31\ X-ray transients with \swift,
\citet[][]{2008A&A...489..707V} identified a SSS transient with the nova\linebreak
M31N~2006-11a. It is detected both in X-rays and the UV about half a year after
the optical maximum and decayed below the \swift\ detection threshold within a
month.

In the ``Deep \xmm\ Survey of \m31" catalogue 40 SSS were detected. Fourteen of
them can be classified as optical nova counterparts by correlations with the  
\m31\ nova catalogue (see below). While many of the sources are already in the
list of PFF2005, four sources are reported for the first time. They were
serendipitously detected 299~d -- 1167~d after the optical outburst (see Stiele
et al., this issue). 

\section{The M 31 nova catalogue} \label{optnova}
About 60\% of the $\sim$65 novae 
occurring in \m31\ per year are located in the central regions, with a
rate in the bulge of  $38^{+15}_{-12}~yr^{-1}$ 
\citep[][]{2006MNRAS.369..257D}.
The central field of \m31\ was continuously monitored since 1997 initially
with WeCAPP
and later using telescopes
at Skinakas (Greece), SuperLOTIS (Kitt Peak) and several other sites 
(Fig.~\ref{fovima}). 
Novae are amongst the brightest variable sources in the data set 
and, due to the
rather dense time coverage, time of outburst and speed class can be well 
determined.  
The optical monitoring is essential for later identification of SSS counterparts. 
In addition to \m31\ optical nova searches at other sites, the optical monitoring of 
\m31\ by our collaboration will be carried on   
during 2009/10 securing close coverage for nova detections also for the 
latest accepted X-ray observations. 
After the commissioning phase, the PanSTARRS1 project will monitor daily 7 deg$^2$ of 
\m31\ for variable objects starting in summer 2009 (PAndromeda). 
GALEX will monitor daily the same field for several weeks in September/October 
2009 in the near and far UV.
These programs are supplemented by a H$\alpha$ nova search program
that allows us, due to the longer detectability of novae 
in H$\alpha$, to identify novae missed during observation gaps, and 
by a program for \m31\ nova spectroscopy at SAO RAS 6m, Skinakas (Greece) and
Asiago (Italy) to determine their type (Fe~II, He/N). Additional spectroscopy is
assured by fast publication of newly detected novae in ``The Astronomer's
Telegram" and direct notification of interested observers. 

We combined the WeCAPP nova list with novae from other 
surveys of \m31. Many of the new detections are listed in the nova pages 
``M31 (Apparent) Novae Page"
provided by the International Astronomical Union, Central Bureau for 
Astronomical Telegrams CBAT\footnote{http://www.cfa.harvard.edu/iau/CBAT\_M31.html} 
and the finding
charts and information, collected by David Bishop\footnote{http://www.rochesterastronomy.org/novae.html}. We 
also\linebreak adopted the CBAT naming scheme for novae that were not registered by CBAT.
We intend to update the internet pages regularly and encourage observers
to provide input for historical and 
forthcoming optical novae in \m31. We will
include photometric and spectroscopic data of optical novae and candidates in 
\m31\ covering all wavelengths. The table is available on our \m31\ nova web
page\footnote{http://www.mpe.mpg.de/$\sim$m31novae/opt/m31/index.php}. Similar
catalogues have been compiled for novae in other Local Group galaxies and are
available from our Local Group nova web
pages\footnote{http://www.mpe.mpg.de/$\sim$m31novae/opt/index.php}.

\section{Position improvement and search for novae on archival plates}
During the catalogue work we noticed that the positional accuracy for many of the 
old novae is poor. This prevents the secure identification of recurrent novae
specifically in the \m31\ center where the nova density is high. 
We therefore initiated a re-analysis of the Tautenburg plates of \m31.
The analysis of the digitized plates yielded 22 new nova candidates and improved positions for 84 novae
\citep[][]{2008A&A...477...67H}. In collaboration with M. Orio we are working to
improve positions of the novae of the ``Rosino et al." catalogues by analyzing 
digitized original plates. It would be important
to also re-analyze other archival plates (e.g. from Hubble, Baade, Arp, Sharov 
\& Alksnis).

\section{Ongoing \xmm/Chandra monitoring of the \m31\ center}
Following the serendipitous detection of SSS phases of optical novae in \m31\
by PFF2005 (see Sect.~\ref{Xnova}) we proposed in 2005 -- and got approved -- a dedicated 
\xmm/\linebreak\chandra\ monitoring program of the \m31\ center area consisting of four 
\xmm\ and four \chandra\ observations separated by 1.5 months. In this way we 
hoped to characterize the SSS
light curves of several novae. Henze et al. (this issue) give some results from
these observations. 

Working on the PHS2007 paper we noticed that by using  
monitoring observations with 1.5 months spacing, 
one still would not be able to follow -- or one even could miss -- the very 
interesting very short SSS phases of novae involving massive WDs. In 2006, we 
therefore proposed a monitoring strategy that would not cover the entire year
but would allow us to
specifically monitor these short duration SSS states. We were granted ten
\xmm/\chandra\ observations with 10 day spacing starting in November 2007. The
optical window for efficient \m31\ nova monitoring opened well before that (June
to February) securing an optical nova catalogue for detecting fast SSS states as 
complete as possible. 

We detected nine
novae in X-rays, four within 4 months after the optical outburst (see
Fig.~\ref{fovima}). The SSS state of three novae lasted less than 3 months with 
M31N 2007-11a holding the record of just 60 days. Following 
\citet[][]{2006ApJS..167...59H}
only novae with WDs with very high masses\linebreak 
($>1.3$~M$_{\odot}$) are expected to show such a short SSS phase 
\citep[][and Henze et al., this issue]{2009A&A...498L..13H}. 
One SSS
correlated with the first nova detected in a globular cluster 
\citep[the He/N nova M31N 2007-06b in Bol 111, see][and Henze et al., this
issue]{2007ApJ...671L.121S,2007ATel.1294....1P,2009A&A...500..769H}.
We even detected a second SSS in a globular cluster
\citep[][and Henze et al., this issue]{2007ATel.1296....1H,2007ATel.1306....1H,2009A&A...500..769H}. 
SSS and novae in globular clusters
are extremely rare objects, just one SSS and two novae were known in globular
clusters before the \m31\
detections. \citet[][]{2009A&A...500..769H} discussed the
properties of these SSSs, unsuccessfully 
searched for a nova counterpart of the second SSS in a \m31\ globular cluster and estimated 
the nova rate in \m31\ globular cluster systems.
Our monitoring data for the interesting SSS counterpart of nova M31N 2007-12b are 
discussed below.

These first results showed that the new monitoring strategy allows us
to effectively select
peculiar cases of SSSs and novae and to obtain a better understanding of the SSS 
nova population. This monitoring is also essential to improve the poor statistics of SSS
and novae in globular clusters. An extension of the program was granted for 
2008/2009 and 2009/2010. 

\section{Periodicity in the SSS counterpart to nova M31N 2007-12b}
\begin{figure}
\includegraphics[width=82mm,clip=]{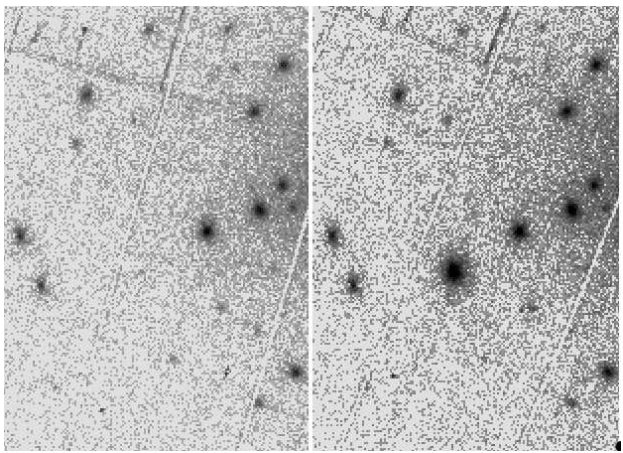}
\caption{\xmm\ EPIC PN images (0.2--1~keV) of the M31N 2007-12b
region, obtained on 2007 December 29 (left) and 2008 January 8 (right).}
\label{img:nova}
\end{figure}
\begin{figure}
\includegraphics[width=65mm,angle=-90,clip=]{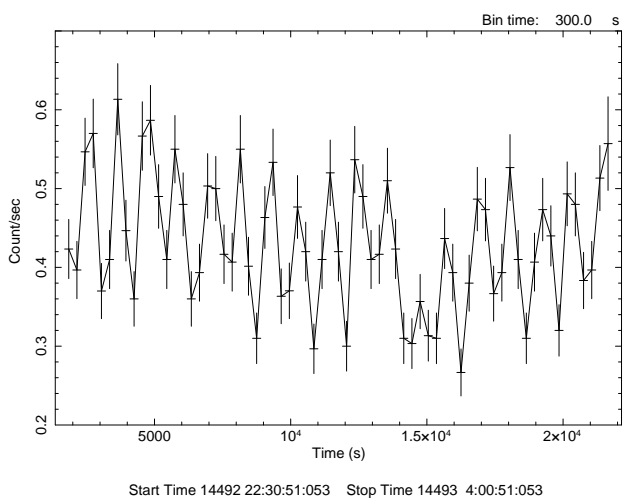}
\caption{\xmm\ EPIC PN light curve (0.2--1~keV) of M31N 2007-12b
 obtained on 2008 January 18.}
\label{lc:nova}
\end{figure}

The optical outburst of nova \n12b\ was detected by several groups 
\citep[see entries on the Central Bureau for Astronomical Telegrams CBAT M31
(Apparent) Novae Page\footnote{http://www.cfa.harvard.edu/iau/CBAT\_M31.html\#2007-12b} and][]
{2007ATel.1324....1L}. The time of outburst can be constrained to less than one 
day by the report from Nishiyama and\linebreak Kabashima (first detection with 16.1 mag 
and last non-\linebreak detection with lower limit of 18.9 mag on 2007 December 9.528 
and 8.574 UT, respectively). From optical spectroscopic data obtained on 
2007 December 15, the nova is classified as He/N
with a Full Width at Half Maximum
of the H$\alpha$ line of $\sim$4500 km s$^{-1}$ 
\citep[][]{2007ATel.1332....1S}.
\citet[][]{2008ATel.1360....1K} reported the discovery of a new X-ray source 
with Swift on 2008 January 13 at a position consistent with the position of 
the optical nova \n12b. No
source was present at the position in Swift observations on 2007 December 16 
and December 30. The emission of the new source was supersoft and the source 
interpreted as a supersoft transient
associated with \n12b\ detected 34~d after the optical outburst.

In our monitoring program no counterpart of nova\linebreak M31N~2007-12b was detected
in \chandra\ HRC-I observations on 2007 December 7 and 17.
In \xmm\ observations on December 29 (21~d after the optical outburst) a very 
faint source is detected which increased in brightness by a factor of more than
200 (\xmm\ observation on 2008 January 8, see Fig.~\ref{img:nova}). In three 
consecutive \xmm\ observations (2008 January 18, 28, and February 7 until 
60~d after the optical outburst) the source stayed bright and showed a supersoft
spectrum. During these observations,
1105s pulsations are detected (see e.g. the light curve of January 18,
Fig.~\ref{lc:nova}). The pulsation period stayed constant (within the errors) 
during the four observations separated by 30~d which suggest this period as the
WD rotation period. During three observations ``dips" can be identified which
may be explained by scattering material within the orbit. These results may
indicate that the nova outburst occurred in a cataclysmic variable system with 
a magnetic WD. Pulsations in \m31\ SSS have been reported for two other systems.
\citet[][]{2001A&A...378..800O} reported 865~s pulsations from a transient
source which may also have been detected during the SSS state of an optical
nova. The second source is the brightest persistent SSS in \m31\ in which 
\citet[][]{2008ApJ...676.1218T} detected 217~s pulsations.

A detailed analysis of the \n12b\ X-ray counterpart is in progress ({Pietsch et 
al., in preparation}). 

\section{Optical nova statistics and outlook}
\begin{figure}
     \includegraphics[width=58mm,angle=-90,clip]{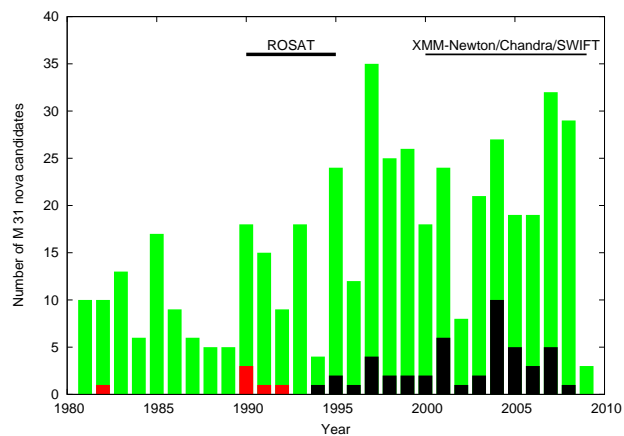}
     \caption[]{Histogram of the number of optical novae detected per year in \m31\ as
     contained in the nova catalogue used for X-ray cross-correlation (see 
     Sect.~\ref{optnova}). The number of optical novae showing X-ray emission 
     is indicated coded for ROSAT and \xmm/\chandra/\swift. The time span of X-ray
     observations is indicated.
     }
    \label{nova_year} 
\end{figure}
Figure~\ref{nova_year} shows the number of optical novae detected per year 
as collected in the
optical nova  catalogue of \m31\ used for cross-correlation with the X-ray
data.  X-ray detected  novae are indicated separating ROSAT and
\xmm, \chandra, or \swift\ detections.  The time span  of the \m31\ observations of
these satellites is also indicated.  Only few optical novae were detected
in the years before the ROSAT observations,  which may explain the lack of nova
identifications with ROSAT SSSs. With \xmm\  and \chandra\  several novae
were detected which had their optical outburst years before the detection of 
X-ray outburst. This may
be caused by very long supersoft states of these novae.
Many novae with short supersoft
states (shorter than 6 months) may have been missed in the sparse and inhomogeneous 
sampling of the light curves and even during the denser monitoring for the center area
starting in November 2007. The crowding of X-ray sources in the very center may
also prevent the detection of novae in \xmm\ observations.

\begin{table}
\caption{Optical nova statistics in Local Group galaxies extracted 
from our Local Group nova web pages http://www.mpe.mpg.de/$\sim$m31novae/opt/index.php, for the Galaxy 
see http://www.cfa.harvard.edu/iau/nova\_list.html.}
\label{tstat}
\begin{tabular}{lrrr}\hline
galaxy & opt. novae & novae/year & detected\\ 
       &      total & av.(2000-2008) & as SSS\\
\hline
Galaxy & 391 &   $\sim$8 & $\sim$10\\
LMC    &  39 & 0.7 &  3\\
SMC    &  17 & 0.3 &   \\
\\
M 31   & 816 &  22 & 53\\
M 32   &   3 & 0.2 &   \\
NGC 205 &  4 & 0.1 &   \\
\\
M 33   & 32  & 0.5 &  2\\
\hline
\end{tabular}
\end{table}
In Table~\ref{tstat} we compare detection rates of optical novae and SSS
counterparts in Local Group galaxies. \m31\ by far leads the list for detected
novae per year and for detected SSS counterparts. This demonstrates the
importance of \m31\ optical nova detections and spectral classifications as well
as X-ray monitoring to get a better understanding for
the percentage of optical novae showing SSS states and statistics on the
dependence on nova type. Only \m31\ allows us to determine the duration of many 
optical nova SSS states to constrain ejected, burning, and WD masses. 
While deep observations of individual novae in the Galaxy allow detailed
investigations using high time and spectral resolution, the large number of
objects in \m31\ allows for a higher probability to find rare objects. This has
been proven by the diversity of nova counterparts detected 
from the monitoring of the center of \m31\ during the last few years.

\acknowledgements
Based on photographic data of the National Geographic Society -- Palomar
        Observatory Sky Survey (NGS-POSS) obtained using the Oschin Telescope on
        Palomar Mountain.  The NGS-POSS was funded by a grant from the National 
        Geographic Society to the California Institute of Technology.  The      
        plates were processed into the present compressed digital form with     
        their permission.  The Digitized Sky Survey was produced at the Space   
        Telescope Science Institute under US Government grant NAG W-2166.
I am grateful to all the members of the \xmm\ \m31\ large program team and the
\xmm/\chandra\ \m31\ nova monitoring collaboration. 

\bibliographystyle{aa}
\bibliography{./novae,/home/wnp/data1/papers/my1990,/home/wnp/data1/papers/my2000,/home/wnp/data1/papers/my2001,/home/wnp/data1/papers/my2004,/home/wnp/data1/papers/my2006}

\end{document}